\documentclass[aps,amsmath,amssymb,preprintnumbers,groupedaddress,twocolumn]{revtex4}
\usepackage{graphicx}

\newcommand{\nc}{\newcommand}
\nc{\beq}{\begin{equation}} \nc{\eeq}{\end{equation}} \nc{\bea}{\begin{eqnarray}}
\nc{\eea}{\end{eqnarray}}

\def\gsim{\mathrel{\rlap{\lower4pt\hbox{\hskip1pt$\sim$}}
    \raise1pt\hbox{$>$}}}       

\def\K3{{\bf K3}}

\def\ov{\overline}

\def\n2d{\cN_{V^*}^{\otimes 2}}

\def\cN{{\mathcal N}}

\def\to{\rightarrow}

\begin{document}

\preprint{UPR-1198-T}
\title{A string theoretic model of gauge mediated supersymmetry beaking}

\author{Mirjam Cveti{\v c}} 
\author{Timo Weigand}
\affiliation{Department of Physics and Astronomy, University of Pennsylvania, Philadelphia, USA}

\begin{abstract}
\noindent
 We propose  a robust supergravity  model of dynamical supersymmetry breaking and gauge mediation, and   a natural embedding in non-perturbative string theory with D-branes. A chiral field (and  its mirror) charged under ``anomalous'' U(1)'s acts as a Polonyi field whose  hierarchical Polonyi-term can be generated by
  string instantons. Further quartic  superpotential terms  arise
 naturally  as a tree-level decoupling effect of
    massive string states.
A robust  supersymmetry breaking minimum allows for gauge mediation with soft masses at the TeV scale,  which we  realise
  for  a   globally consistent SU(5) GUT model of Type I  string theory, with a D1-instanton inducing the Polonyi term.

\end{abstract}


\maketitle

\date{today}

\bigskip


{\bf I. Introduction} One of the mysteries of particle physics is if Nature has chosen supersymmetry to protect the electroweak scale in the TeV regime and, if so, how
the breaking of supersymmetry  is communicated to the Standard Model.
Among the different mediation mechanisms, gauge mediation \cite{Giudice:1998bp} offers a clean rationale for the absence of
flavour-changing neutral currents. Its   model-independent experimental signatures \cite{Meade:2008wd,Distler:2008bt} make it an interesting
scenario also in light of future testability in the Large
Hadron Collider (LHC) era.
In its simplest implementation, supersymmetry is broken by the F-term of a hidden gauge sector chiral superfield $S$ and communicated to the visible sector by a vector pair of messenger
fields $q,\, {\tilde q}$. Gaugino  and slepton
masses arise at the one- and two-loop level, respectively,
roughly of  order
 $\frac{\alpha}{4 \pi} \, \frac{F}{S}$, while the gravitino mass is of order $F/M_{Pl}$ ($M_{pl}$-Planck mass). One of the challenges in realising gauge mediation is therefore to dynamically generate
a sufficiently small vacuum expectation value (VEV) for the supersymmetry breaking field $\langle S \rangle < 10^{-3} \, M_{Pl}$
in order for the  effects of gravity mediation
  to be subleading \footnote{
For an early discussion within  string theory see
\cite{Diaconescu:2005pc}.}.

The perhaps simplest supersymmetry breaking scenario
involves a linear superpotential of Polonyi
type $\mu^2 S$ whose F-term  breaks
supersymmetry at the scale $\mu^2$.
In string theory,  D-brane  instantons
 \cite{Blumenhagen:2006xt,Ibanez:2006da,Florea:2006si}
can account  both for the presence of the Polonyi term \cite{Aharony:2007db} and for a hierarchical suppression of its scale $\mu$,  as demonstrated  even in globally consistent examples \cite{Cvetic:2007qj}.
Yet, in order to realise gauge mediation, further dynamical input is required to stabilise $S$ at the desired hierarchical scale.
This can be achieved  \cite{Kitano:2006wz} by including
one-loop K\"ahler potential corrections with a specific sign \footnote{For a recent alternative approach see e.g. \cite{Murayama:2007fe},
and for a stringy mediation mechanism using D-instantons \cite{Buican:2008qe}.}.

In this article we propose an alternative model where the Polonyi field is stabilised
with the help of quartic superpotential terms.
These generically result from integrating out heavy string states provided the Polonyi field is charged under
massive $U(1)$ gauge factor(s).
The specific terms we are interested in  require a vector pair $\tilde S$ with
 opposite $U(1)$  charge(s).
Models of this type arise naturally in string theory
 with D-branes.
Under specific assumptions on the moduli dynamics this framework possesses a supersymmetry breaking vacuum tailor-made for gauge mediation to generate TeV scale soft masses. Within   Type I  compactifications with D-instantons we also present a globally consistent Grand Unified Theory (GUT) model  of the  type  discussed in \cite{Cvetic:2007qj} where the stringy consistency conditions allow for TeV soft masses.
\vspace{5pt}

{\bf II. The model} Our supersymmetry breaking hidden sector consists of a massive $U(1)_a \times U(1)_b$ gauge theory with bi-fundamental chiral superfields $S_{(-1_a,1_b)}$ and $\tilde S_{(1_a,-1_b)}$ and
\bea
\label{model}
&& W=\mu^2S+c-\frac{S^2 \tilde S^2}{4M} - \lambda S q \tilde q, \\
&& K=SS^\dagger +{\tilde S}{\tilde S}^\dagger + q q^\dagger+ {\tilde q}{\tilde q}^\dagger. \label{WK}
\eea
The messenger fields $q_{-1_b}$, $\tilde q_{1_a}$ form a vector-like pair under the visible sector gauge group and will be responsible for gauge mediation of supersymmetry breaking.
As will be detailed in the next section  the above model is well-motivated from string theory:
The massless vector-like pair $S$ and $\tilde S$ arises for instance at the intersection of a pair of D6-branes in the Type IIA context. The gauge bosons of $U(1)_a$ and $U(1)_b$ acquire string scale masses via the Green-Schwarz mechanism. The perturbatively forbidden Polonyi term $\mu^2 S$ for a charged field can naturally be generated by D-brane instantons \cite{Cvetic:2007qj}\footnote{The same instanton may also generate the terms of the type $S^2{\tilde S}$ whose coupling is of the order of $\mu^2/M_{s}^2$ ($M_s$-string scale) and thus for small values of $S$ and ${\tilde S}$ it is negligible compared to the Polonyi term. We thank E. Dudas for a discussion on this point.}.
Gauge invariance is ensured by the shift of the axionic part of the closed modulus $T$ which determines the part of the instanton action charged under the massive $U(1)$s \cite{Blumenhagen:2006xt,Ibanez:2006da,Florea:2006si} in $\mu^2 = \mu_0^2\, e^{-T}$.
Higher monomials in $S$ or $\tilde S$ only are perturbatively absent due to charge selection rules.
The constant $c$ results from integrating out the closed string moduli uncharged under $U(1)_a $ and $U(1)_b$, which are assumed to be stabilised at a scale much higher than $\mu$.

The non-renormalisable quartic term in the superpotential arises from \cite{Cvetic:1998ef} the  integration of massive (closed or suitable open string sector) modes $C$ with mass $M_C$ which couple in the superpotential  to  $S{\tilde S}$ as
\bea
\label{int1}
W_{C}=  {\lambda}_C \, C S \tilde S +{M_C} \, C^2  \, ,
\eea
with $M={M_C}/{\lambda_C^2}$.
 For heavy closed string states  and $\lambda_C\sim 1$, $M \sim  M_s$ -- the string scale. By contrast, if $M_C$ is associated with
  dynamically generated masses for (closed and open sector) string moduli,
   $M$ could be $\ll M_{s}$.  The effective $M$ can in principle decrease significantly due to enhanced threshold effects of higher mass level string states $C$ whose multiplicity increases exponentially. On the other hand if there were a selection rule that would set $\lambda_C= 0$, the effective $M\to \infty$ and the massive states would decouple.  We shall see that our
    results are extremely robust in that they depend only mildly on
    the values of $M$  and are  basically driven
    by $\mu\ll 1$  in the interplay between the Polonyi
    and the tree-level quartic superpotential term.

The  superpotential  $W_C$
(\ref{int1})
 also induces tree-level K\"ahler potential
corrections \cite{Cvetic:1998ef}
$\delta K_1 = +\frac{ SS^\dagger {\tilde S}{\tilde S}^\dagger}{4\,M^2}\, .$
$K$ can furthermore receive one-loop corrections due to the couplings of $S$ and ${\tilde S}$ to bi-linears of open-string heavy states with mass ${\tilde M}$,
$\delta K_2\sim -\frac{(SS^\dagger)^2}{\Lambda^2}-
 \frac{({\tilde S}{\tilde S}^\dagger)^2}{{\tilde \Lambda}^2}\, ,$
where $\Lambda\sim {\tilde \Lambda}\sim \pi {\tilde M}$.
As we shall see momentarily, for our explicit solution $S \sim
{\tilde S}\ll 1$
both $\delta K_1$ and $\delta K_2$ can be neglected relative to the superpotential  term $-\frac{\,S^2{\tilde S^2}}{4M}$.

The full scalar potential
$V=V_F + V_D$ takes the standard ${\cal N}=1$ supergravity form
\bea
\label{POT}
 V_F&=& e^{K} (D_i W \, D_{\ov j} W K^{i \ov j}- 3 |W|^2), \nonumber \\
 V_D&=& {\textstyle{\frac{{g_a^2}}{2}}}(-|S|^2 + |\tilde S|^2+|q|^2+\xi_a)^2 + \\
 &&     {\textstyle{\frac{{g_b^2}}{2}}}(|S|^2 - |\tilde S|^2-|{\tilde q}|^2+\xi_b)^2\, \nonumber \label{Vtot}
\eea
in terms of $D_i W= \partial_i W + K_{,i} \, W$ and the gauge couplings $g_{a,b}$ associated with $U(1)_a$ and  $U(1)_b$, respectively. We take
$M_{Pl}=1$.
The Fayet-Iliopoulos (FI) terms $\xi_a$ and $\xi_b$ depend on $T$ and in 
general also on other  closed moduli $N_i$ not entering the superpotential (\ref{model}).
Their full dynamics hinges upon the precise form of their K\"ahler potential \cite{Cicoli:2008va}. To avoid such model dependent questions we do not analyse the stabilisation of $T$ and $N_i$ explicitly here but assume they can be stabilised such that the FI terms are at most of order  $\mu^{8/3}$  (see later); furthermore under specific conditions on the K\"ahler potential for $T$ and $N_i$ their backreaction on the effective potential (\ref{POT}) is  negligible. An analysis of these conditions will be presented elsewhere.

Under these assumptions on the moduli sector there exists a stable solution in the regime $\langle S \rangle  \sim \langle {\tilde S}\rangle = {\cal O}(\mu^2\, M)^{\frac{1}{3}}) \ll 1$
and
 $\langle q\rangle =
\langle {\tilde q} \rangle=0$, as  ensured by their F-terms.
We reparametrise $S$ and $\tilde S$ as $S=|S|\exp(i\phi)$, ${\tilde S}=|{\tilde S}|\exp(i{\tilde \phi})$ and proceed iteratively by first enforcing vanishing D-terms via
$ \langle |S|\rangle =\langle |{\tilde S}|\rangle  \equiv s\,$, neglecting the FI terms at this stage.
It is convenient to introduce a new combination of fields $S_{\pm}=(|S|\pm |{\tilde S}|)/\sqrt{2}$, where $S_-$ obtains the dominant mass $m^2_{S_-}=4(g_a^2+g_b^2)s^2\,$ from the D-term.
The supergravity potential  $V_F$  in (\ref{Vtot}) is in this approximation expanded only in terms of $(S_+,\,  \phi, \, \ {\tilde \phi})$. Since we are looking for a minimum  in the  region
$\langle S_+\rangle = {\cal O}((\mu^2M)^{\frac{1}{3}})\ll 1$
it suffices to expand only up to terms  proportional to  $\mu^4 \langle S_+\rangle = \mu^4\sqrt{2}s$.
The leading order potential then takes the form
\begin{equation}
V_{F_0}= \mu^4-3{c^2}-\frac{\mu^2 s^3}{ M}\cos(\sqrt{5}\phi_1)+\frac{s^6}{2M^2}
\, ,\label{VF0}\end{equation}
where we have introduced new fields
$\phi_1\equiv ({\phi +2{\tilde \phi}})/{\sqrt{5}}  \, , $ and $   \phi_2\equiv (-2 \phi +{\tilde \phi})/{\sqrt{5}}\, .$
Note that up to the term $-3 c^2$ this is the potential obtained in the globally supersymmetric approximation.
The minimum of  ({\ref{VF0}) and its zero value there are  ensured  by taking  respectively
\begin{equation}
 \phi_1=0\, , \ \  s=\left({\mu^2 M}\right)^{\frac{1}{3}} \, ;\ \
c={\mu^2}/{\sqrt{6}}\, ,
\label{cs}\end{equation}
 while
 $\phi_2$  has a flat direction.
In the next step we correct for
having set $\langle S_-\rangle  = 0$ in $V_{F_0}$ and find
$\langle S_-\rangle ={\cal O}(\mu^2/((g_a^2+g_b^2)\, M)) \ll \langle S_+\rangle$. The tiny deviation from $S_-=0$ entails a subleading $D$-term of ${\cal O}(\mu^{\frac{8}{3}}/M^{\frac{2}{3}}) \ll F$. The backreaction from the closed string moduli dynamics is likewise subleading under the above assumption of FI terms scaling at most like $ \mu^{8/3}$. This  justifies the iterative procedure and  the correction to $\langle S_+\rangle$ is negligible.
At this order the potential respects R-symmetry and  $\phi_2$ is the Goldstone boson of this spontaneously broken global symmetry.
The degeneracy of $\phi_2$ is removed due to supergravity effects
 in the potential at the next order, linear  in powers of $s$
 (for simplicity we set $\phi_1=0$ at this order),
\begin{equation}
V_{F_1}=-\frac{ c s \mu^2}{2} (8+\frac{s^3}{\mu^2 M}) \cos(\frac{2\phi_2}{\sqrt{5}})\, ,
\end{equation}
which fixes
$\phi_2=0\,$.
At this order $c$ and $s$ are likewise corrected, but the corrections are suppressed by ${\cal O}(s)\ll 1$ and thus again
subleading.
  The scalar $S_-$ is much heavier than   $S_+$ and $\phi_1$, while
 $\phi_2$ has a  positive mass-square
  for positive  $c$ (at this level)  and  is  further suppressed by  one power of $s$ 
   For the values  (\ref{cs}) one obtains
\begin{equation}
m^2_{S_-}=4(g_a^2+g_b^2)M^{\frac{2}{3}}\mu^{\frac{4}{3}}\, , 
m^2_{S_+}={\textstyle{\frac{9}{4}}}\, M^{-\frac{2}{3}} \mu^{\frac{8}{3}}
={\textstyle{\frac{9}{10}}}
m^2_{\phi_1}\, ,
\nonumber\end{equation}
and
\begin{equation}
m^2_{\phi_2}=\textstyle{\frac{9}{5}}\, c \, M^{-\frac{1}{3}}\mu^{\frac{4}{3}} = {\textstyle{\frac{3\sqrt{6}}{10}}}
M^{-\frac{1}{3}}\mu^{\frac{10}{3}}\, .
\end{equation}
The model predicts  ${3}/{\sqrt{10}}$ for the mass ratio
 of $S_+$ and $\phi_1$. Note that only $m^2_{\phi_2}$ depends on the value of $c$.
The mass-square correction $\delta_D m^2_{S+}$ due to the small D-term is of $ {\cal O}(\mu^4/M^2)$ and thus subleading.

The F-term  messenger masses are of the order $\lambda s$
and positive as long as $\mu^2 \le \lambda s^2$ or equivalently $\mu\le \lambda^{\frac{3}{2}}M$. This is  satisfied for  the relevant range  of $\mu \sim 10^{-10}$ (see below) and
  $\lambda\gg 10^{-6}$.  The D-term mass corrections $\delta_D m_{q} = {{\cal O}(\lambda s \, \mu^{\frac{2}{3}}/M^{\frac{2}{3}})}$ are again subleading. The model has a small gravitino mass $m_{3/2} =\mu^2$.

\noindent {\it Coleman-Weinberg one-loop corrections}\ \
 due to the  superpotential coupling  of $S$  to the messengers $q$ and ${\tilde q}$  in (\ref{WK})  result in the K\"ahler potential correction
\begin{equation}
\delta K= -\kappa SS^\dagger \log ({\textstyle{\frac{SS^\dagger}{\Lambda^2}}})\, , \ \ \ \ \kappa\equiv \frac{\lambda^2N_c}{16\pi^2}\,,
\end{equation}
 where the renormalisation scale $\Lambda$ at which the coupling $\lambda$ is defined  is chosen to be of the order of the VEV of $S$.  $N_c$ is the number of colors associated with the observable sector  gauge group $SU(N_c)$, e.g.,  $N_c=5$  for $SU(5)$ GUT, and the messengers are in respective ${\bf N_c}$ and ${\bf {\overline N_c}}$ representations.
 Since these corrections  respect $R$ symmetry they cannot modify the mass of $\phi_2$ and their contribution to the masses of other scalars are only subleading for $\lambda \le 1$ \footnote{
Other one-loop corrections due to the  $(S, \ {\tilde S})$ sector lead to K\"ahler potential corrections of the type $\sim \frac{(SS^\dagger)^2}{M^2}\log(\frac{SS^\dagger}{\Lambda^2})$ and are thus subleading. A deviation from D-flatness,
due to a small non-zero $S_-$, leads to negligible one-loop corrections
proportional to $S_-^2S_+^2$.
}.


\noindent{\it Phenomenological Analysis}\ \
Due to the relative suppression of the D- versus the F-term, the supersymmetry soft masses are dominated by gauge-mediated F-term breaking.
The loop-generated visible sector soft  masses  are determined by  $m_{soft}\sim \frac{\alpha}{4 \pi}\, {\langle F\rangle }/{\langle
 S \rangle } \sim 10^{-3} {\mu^2}/{s}$ \cite{Giudice:1998bp} and lie in the
 TeV range  provided  $\mu^2\sim 10^{-13} s$;
 in this case
 the solution ({\ref{cs}) for $s$ implies a relationship
 $\mu\sim  10^{-10}\ M^{\frac{1}{4}}\, $ and consequently
  $s\sim 10^{-7}M^{\frac{1}{2}}.$
The corresponding hidden sector scalar masses are predicted to be in the range
\begin{eqnarray}
m_{S_-} \sim
10^{10}\, -\, 10^{11}\,\hbox{GeV} ,\
\  && m_{S_+}\sim
10^{3}\, -\, 10^4\,
\hbox{TeV} , \nonumber\\
m_{\phi_1}
\sim  10^{3}\, -\, 10^4\, \hbox{TeV} ,  \ \
&&m_{\phi_2}\sim
1\, -\, 10^{2} \, \hbox{TeV}\, .
\nonumber \end{eqnarray}
    The messenger masses are in the range  $10^{10}\, -\, 10^{11}$ GeV. Interestingly, the model has a light  gravitino of mass in the range $ 0.1 \, -\,  10 $ GeV.
In table I we  present numerical values for a wider parameter range of $\mu$  and $M$.

\begin{table}[ht]
\centering
\begin{tabular}{|c|c|c|c|c|c|c|c|}
\hline
 $\mu$  & $M$ &
  & $m_{S_-}$& $m_{S_+}$ & $m_{\phi_2}$ & $m_q$  &$m_{3/2}$  \\
\hline \hline
$10^{9}$&$10^{18}$&
&$2.83\, 10^{11}$ &$1.50\, 10^{6}$ &$2.71\, 10^{2}$ &
 $1.00\, 10^{11}$&  $0.10$\\
 $10^{9}$&$10^{16}$&
 &$6.08\, 10^{10}$ &$6.96\, 10^{6}$ &$5.84\, 10^{2}$ &$2.15\, 10^{10}$ & $0.10$ \\
$10^{10}$&$10^{18}$&
&$1.31\, 10^{12}$ &$3.23\, 10^{7}$ &$
1.26\, 10^{4}$ &$4.64\, 10^{11}$ & $10.0$\\
$10^{10}$&$10^{16}$&
&$2.83\, 10^{11}$ &$1.50\, 10^{8}$ &$2.71\, 10^{4}$
&$1.00\, 10^{11}$ &  $10.0$\\
\hline
\end{tabular}
\caption{
Masses for
$S_{\pm}$  and $\phi_2$,
  messengers $q$ and the gravitino for different values of
 $\mu$ and  $M$. Note, $m_{\phi_1}={\sqrt{10}}
 m_{S_+}/3$ and $m_{\tilde q}= m_q$. We took $\lambda =0.10$, $g_a=g_b=0.10$, $N_c=5$
 and $M_{Pl}=1.22\, 10^{19}$ GeV.  All masses are in GeV.
\label{TI} } 
\end{table}
The above analytic results are
 extremely close to actual numerical values of the potential:
corrections to these analytic expressions, which would appear in the potential at the  $\mu^4 s^2$ order, modify the above
expressions at a level of $100\times{\cal O}(s) \, \% \sim 10^{-5}\% $.
 Note also that  K\"ahler potential corrections
 due to massive modes, as described above,  contribute only
 at this order. Therefore even if we lower the
 string scale $M$ and  ${\tilde \Lambda}$ to,
 say, $10^{-3}$  and  $ 10^{-2}$, respectively,
 these corrections are small.

\noindent{  \it Other Minima} \ \
 There exists no nearby supersymmetric minimum  with non-zero VEV for the messenger fields, unless there are additional fields $(q^\prime_{1_b},{\tilde q}^\prime_{-1_a})$
to ensure D-flatness as $s\to 0$ and  $\langle q\rangle =\langle {\tilde q}\rangle\to \mu/\sqrt{\lambda}$. This would lead to
a supersymmetric vacuum with vacuum energy $-\mu^4/2$.  Our solution is stable against false vacuum decay into this supersymmetric one as the  bounce action can be estimated to be  $ \sim \pi^2(\Delta s)^4/\Delta V \sim  \pi^2 \mu^{8/3}/\mu^4\sim 10^{30}$.   In absence of mirror fields $(q^\prime, {\tilde q}^\prime)$, there exists only a non-supersymmetric nearby solution  with $\langle S\rangle =\langle q\rangle =\langle {\tilde q}\rangle = \mu/\sqrt{3\lambda}$ and $\langle {\tilde S}\rangle =0$ and positive energy   $+\mu^4/6$.
Note also that as  $s\to {\cal O}(1)$ the K\"ahler potential corrections, discussed above, lead to a singular K\"ahler metric and  the blow-up of the potential.

{\bf III. Global string realisation} The described model of gauge mediation has a natural realisation in string theory. For definiteness our discussion focuses on
intersecting D-brane models in Type IIA or Type IIB Calabi-Yau orientifolds where
massless charged matter arises from strings stretching between different D-branes. Generalisations to the respective strong coupling M- or F-theory versions are likewise possible.

A particularly economic realisation appears in the context of $SU(5)$ GUT models. The hidden sector consists of two stacks of single D-branes wrapping (possibly magnetised) cycles $\Pi_a$, $\Pi_b$ with a massless vector pair $S$ and $\tilde S$ in the $(a,b)$ sector. The SU(5) gauge group can arise from a stack of $5$ coincident branes on $\Pi_c$ with the ${\bf \ov 5}_m$ and $[{\bf 5}_H + \ov {\bf 5}_H ]$ localised at intersections with another single brane $\Pi_d$.
In addition, the setup contains the orientifold image of each brane stack and matter in the, say, $(a,b)$ sector is identified with the image sector and $(b',a')$.
A possible identification of the visible, hidden and messenger sector matter is given in table \ref{TII}.
\begin{table}[ht]
\centering
\begin{tabular}{|c|c|c||c|c|c|}
\hline
 Particle  & Charge & Sector&  Particle  & Charge & Sector  \\
\hline \hline
$(Q_L,U_R^c,e_R^c)$  & ${\bf 10}_{2c}$ & $(c',c)$ & $S$ & $(-1_a,1_b)$ & $(a,b)$ \\
$(L,D_R^c)$ & ${\bf \ov 5}_{(1_d,-1_c)}$ & $(c,d)$ & $\tilde S$ & $(1_a,-1_b)$ & $(b,a)$ \\
Higgs      & $ {\bf 5}_{(1_d,1_c)}$  & $(d',c)$ & $q$ & $ {\bf 5}_{(-1_b,1_c)}  $ & $(b,c)$ \\
$N_R^c$      & $ 1_{-2d}  $  & $(d,d')$ & $\tilde q$ & $ {\bf \ov 5}_{(1_a,-1_c)}  $ & $(c,a)$ \\
\hline
\end{tabular}
\caption{
Embedding of supersymmetry breaking hidden sector into $U(5)_c \times U(1)_d$ GUT theory.
\label{TII} } 
\end{table}

For the sake of applications we now specialise to compactifications of Type I string theory on a Calabi-Yau manifold $X$. The relevant D-branes are space-filling D9-branes carrying holomorphic vector bundles $V_a$. Their orientifold image carries the dual bundle $V_a^{\vee}$.
We will choose the simple case of line bundles $L_a$. The massless open matter in the, say, $ (a,b)$, sector is counted by the cohomology group $H^i(X,L_a^{\vee} \otimes L_b)$, where $i=1$ refers to chiral matter in the bi-fundamental $(\ov N_a,N_b)$ and $i=2$ to the conjugate representation $(N_a, \ov N_b)$. For technical details see \cite{Cvetic:2007qj}.

The Polonyi term $\mu^2 S$ in the superpotential can be generated by Euclidean D1-branes, so-called E1-instantons, which are localised in the four external dimensions and wrap suitable holomorphic 2-cycles $C$ on $X$.
For this to happen the instanton has to carry precisely one  charged fermionic zero mode
$\lambda_a$ and $\lambda_b$ of charge $+1_b$ and $-1_b$, respectively. These arise from massless open strings between the instanton and the two hidden brane stacks with bundles $L_a$ and $L_b$ and are counted by $H^i(C,L_a|_C \otimes \sqrt{K_C})$. Here $i=0$ and $i=1$ refer to chiral modes of charge $1_a$ and $-1_a$, respectively \footnote{Note the slight change in conventions as compared to \cite{Cvetic:2007qj}.}.
The scale of the resulting superpotential is set by the string coupling $g_s$ and the volume of the instanton cycle ${\rm Vol}_C$ in string units as
$W_{np}= M_s^2 \exp(-{\textstyle{ \frac{2\pi}{g_s}}} {\rm Vol}_C )$.

We now demonstrate in an explicit, globally consistent GUT toy model how the scales leading to TeV soft masses in the visible sector can arise. Our main assumption is that the quartic superpotential term is due to massive string modes of string scale
 $M_s$ such that $M= {\cal O}(M_s)$.
In Type I theory, the string scale is determined by the string coupling $g_s$ as
$M_s^2= 2 \pi \, m_g^2\,  g_s \, \alpha_{GUT}$ with $m_g \simeq 2.4 \times \, 10^{18}\,$ GeV. If we assume that $g_s$ is stabilised in the perturbative regime, say,  $g_s=0.4$, then $\alpha_{GUT}=0.04$ implies $M_s=7.6 \times 10^{17}$ GeV and ${\rm Vol}_C$ has to be stabilised around $2.63$ for TeV soft masses.

While the actual stabilisation of $g_s$ and the geometric moduli is not addressed in this paper, we now show in our concrete example that the above sample values are compatible with the non-trivial D-term supersymmetry constraints that arise from the presence of the magnetised D9-branes.
As in \cite{Cvetic:2007qj} we work on an elliptic fibration over $dP_4$. The SU(5) GUT brane $\Pi_c$ and the hidden sector $\Pi_a$, $\Pi_b$ correspond to D9-branes with line bundles $L_c$, $L_a$ and $L_b$ and multiplicities $N$ as in table \ref{TIII}. For details of the notation see \cite{Cvetic:2007qj}. The D5-brane tadpole is cancelled by introducing D5-branes along curves in the effective class $W= 24 F + \pi^*(16l-4E_1-12E_2-4E_3)$, and the K-theory charge $\sum_i N_i c_1(L_i)$ mod 2 vanishes.
 The toy model gives rise to 4 chiral families of ${\bf 10}$ plus additional vector pairs. In contrast to the general setup of table \ref{TII} there is no $U(1)_d$ stack, but the ${\bf \ov 5}_m$ and $[{\bf 5}_H + \ov {\bf 5}_H ]$ arise from the sector between $\Pi_c$ and the D5-branes. From $H^*(L_a^* \otimes L_b) = (0,27,5,0)$ one reads off a considerable excess of hidden sector fields in addition to the minimal $(S, \tilde S)$ pair as in the scenario of the previous section.

The Polonyi term for $S$ arises from an instanton wrapping an isolated ${\mathbb P}^1$ in the divisor $\pi^*E_4$ with zero intersection with the base of the fibration \cite{Cvetic:2007qj}.
\begin{table}[]
\renewcommand{\arraystretch}{1.5}
\begin{center}
\begin{tabular}{|c|c|c|}
\hline \hline
Bundle & N & $c_1(L)= q \sigma + \pi^*(\zeta)$    \\
\hline \hline $L_a$ & 1 & $\pi^*(- l + 2 E_1 + 2 E_2- 2E_3 -E_4) $ \\\hline $L_b$ & 1 & $ 4\sigma +\pi^*(l - 2 E_2 +
 E_4)$ \\\hline $L_c$   & 5 & $ \pi^*(2E_1  -2 E_2 -2E_3)$    \\\hline
\hline
\end{tabular}
\caption{\small A $ U(5)_c \times U(1)_a \times U(1)_b$ Polonyi model.  } \label{TIII}
\end{center}
\end{table}
Indeed we find $H^*(C,L_a|_C \otimes \sqrt{K_C})=(1,0)$ and $H^*(C,L_b|_C \otimes \sqrt{K_C})=(0,1)$ and thus precisely the amount of charged zero modes $\lambda_a$ and $\lambda_b$ required to generate the Polonyi term.
As the instanton cycle is rigid, no subtleties associated with extra uncharged zero modes arise.
 The D-term supersymmetry conditions can be satisfied inside the K\"ahler cone, e.g., for $r_{\sigma}=1.06$, $r_l=9.37$, $r_1=-4.99$, $r_2=r_3=-3.00$, $r_4=-2.63$. The resulting gauge kinetic function $f= \tilde f/(2 \pi g_s)$ with $\tilde f_a=12$ leads to a slightly too small value of $\alpha_{GUT}=g_s/\tilde f_a =0.03$.
The instanton volume is given by $|r_4|=2.63$, as required for $\mu^2 ={\cal O}(10^{-20})$ and TeV soft masses.
While the spectrum of the visible sector is by no means semi-realistic, the example does serve as a prototype demonstrating how our model of gauge mediated supersymmetry breaking can be engineered in string theory. We hope to implement this module into more realistic string vacua.

\emph{Acknowledgements}
We thank G. Shiu for collaboration in initial stages and P. Ouyang and G. Shiu for many important discussions. We
acknowledge useful discussions with J. Marsano, Y. Nomura and R. Richter.
We thank the Aspen Center for Physics for hospitality during the course of this research. T.W. thanks the University of Madison and the Banff Center for hospitality. Our work is supported in part by the DOE Grant DOE-EY-76-02-3071, the NSF RTG DMS Grant 0636606 and  the Fay R. and
Eugene L. Langberg Endowed Chair.




\end{document}